\RequirePackage{fix-cm}
\documentclass[twocolumn]{svjour3}          
\smartqed  
\usepackage{graphicx}
\usepackage{mathptmx}      

\usepackage{latexsym}
\usepackage{tikz}
\usepackage{caption}
\usepackage{subcaption}
\usepackage{tikz}
\usetikzlibrary{spy,calc}
\usepackage{hyperref}
\usepackage{verbatim}
\usepackage{booktabs,array} 
\usepackage{url}
\usepackage{amsmath, amsfonts, amssymb}
\usepackage{siunitx}
\usepackage{microtype}
\usepackage{cite}
\usepackage{textcomp}
\usepackage{breqn}

\usepackage{etoolbox}
\newcommand*{\affaddr}[1]{#1} 
\newcommand*{\affmark}[1][*]{\textsuperscript{#1}}

\usepackage{xcolor}
\def\BibTeX{{\rm B\kern-.05em{\sc i\kern-.025em b}\kern-.08em
    T\kern-.1667em\lower.7ex\hbox{E}\kern-.125emX}}

\usepackage{hyperref}

\usepackage{pgfplots}
\pgfplotsset{compat=newest, legend style={at={(1,0.05)},anchor=south east}
}

\hypersetup{hidelinks,
backref=true,
pagebackref=true,
hyperindex=true,
breaklinks=true,
colorlinks=true, linkcolor=blue,
urlcolor=blue, citecolor=blue,
bookmarks=true}

\newcolumntype{L}[1]{>{\raggedright\arraybackslash}p{#1}}

\makeatletter
\DeclareRobustCommand\onedot{\futurelet\@let@token\@onedot}
\def\@onedot{\ifx\@let@token.\else.\null\fi\xspace}
\makeatother

\makeatletter
\newcommand{\thickhline}{%
    \noalign {\ifnum 0=`}\fi \hrule height 1pt
    \futurelet \reserved@a \@xhline
}

\makeatletter
\renewcommand\footnoterule{%
  \kern-3\p@
  \hrule\@width.4\columnwidth
  \kern2.6\p@}
  \makeatother

\journalname{SN Computer Science}

\begin{document}

\title{Conversion Between Cubic Bezier Curves and Catmull–Rom Splines\thanks{This is the pre-print version. \\ The main article is published in SN Computer Science: \url{https://link.springer.com/article/10.1007/s42979-021-00770-x} \\ DOI: \url{https://doi.org/10.1007/s42979-021-00770-x} \\ Soroosh Tayebi Arasteh and Adam Kalisz have contributed equally to this study. \\ \textbf{Soroosh Tayebi Arasteh} is the corresponding author.}}

\author{Soroosh Tayebi Arasteh\protect\affmark[1,2] \and Adam Kalisz\affmark[1] 
}

\authorrunning{S.T. Arasteh, et al.} 

\institute{\\
              \affaddr{\affmark[1]Department of Electrical, Electronic, and Communication Engineering, Information Technology (LIKE), Friedrich-Alexander-Universit\"at Erlangen-N\"urnberg, Am Wolfsmantel 33, 90158 Erlangen, Germany}\\\\
\affaddr{\affmark[2]Harvard Medical School, 25 Shattuck St, Boston, MA 02115, USA}\\\\
\email{\\
 soroosh.arasteh@fau.de \\ adam.kalisz@fau.de}
}


\maketitle

\begin{abstract}
Splines are one of the main methods of mathematically representing complicated shapes, which have become the primary technique in the fields of Computer Graphics (CG) and Computer-Aided Geometric Design (CAGD) for modeling complex surfaces. Among all, Bézier and Catmull–Rom splines are the most common in the sub-fields of engineering.
In this paper, we focus on conversion between cubic Bézier and Catmull–Rom curve segments, rather than going through their properties.
By deriving the conversion equations, we aim at converting the original set of the control points of either of the Catmull–Rom or Bézier cubic curves to a new set of control points, which corresponds to approximately the same shape as the original curve, when considered as the set of the control points of the other curve. Due to providing simple linear transformations of control points, the method is very simple, efficient, and easy to implement, which is further validated in this paper using some numerical and visual examples.

\end{abstract}

\section{Introduction}

\subsection{Motivation}

One of the main challenges in computer-aided design is finding a suitable shape representation which is both performant and flexible, when implemented in a computer software. Especially with curves, a software engineer is usually presented with a plethora of possible shape representations to choose from. The wide range of candidates and their specific advantages and disadvantages motivate the application of this work. A brief overview of the main purpose of curves and splines as well as the necessity of a conversion method in engineering fields will prepare the reader for the subsequent theory and evaluation part within this paper.

Designers in the car industry often need to construct clay models or sketches of vehicles, where the mathematical representation needs to allow for interactive modifications to curve properties, such as curvature and continuity. Furthermore, one of the key properties of the available mathematical representations is the distinction between \textit{interpolating} and \textit{approximating} curves \cite{101093comjnl1571}. This property is related to whether a curve goes through its \textit{Control Points} or not and can be observed as one of the main differences between the two shape representations discussed in this work. Indeed, the reason for conducting this research was the need to support two different curve representations in a common file format, and therefore, the remainder of this paper is focused on proposing a closed-form conversion equation.

Furthermore, according to the German Association of the Automotive Industry (VDA)\footnote{In German: Verband der Automobilindustrie.}, different manufacturers and their subcontractors have different geometric modeling systems for curve and surface representations. Thus, another motivating view on exchanging data between different geometric modeling systems to compensate differences in the types of polynomial bases, maximum polynomial degrees and mesh sizes of curve and surface representations is discussed more thoroughly in~\cite{HOSCHEK198759}.

\subsection{Engineering Purpose and Related Work}

As the field of Computer Graphics (CG) develops, techniques for modeling complex curves and surfaces are being increasingly important. An example may be the rendering of highly detailed landscapes which need to be subdivided (tessellated) depending on the viewing distance to reduce the processing time for a geometry that is barely visible from the current camera viewpoint \cite{MultilevelTerrainRendering, TesselationSplineSurfaces}. One of the major techniques to realize such dynamic tessellation algorithms is the use of parametric splines in which a curve is defined by piecing together a succession of curve segments, and especially, surfaces defined by stitching together a mosaic of surface patches \cite{BartelsBeatty}.

Splines are a mathematical means of representing a complex curve. For example, it is not possible to define a circle using only a single cubic Bézier curve. Therefore, a common approximation to model a circle is to use four Bézier curve segments~\cite{DOKKEN199033}. Eventually, a complete path can be defined using a series of points at intervals along the curve segments and defining a function that allows to interpolate along the curve to retrieve additional points within each interval to be calculated. This is often the first step when creating procedural virtual worlds, i.e., generating smooth and infinite highways using Catmull–Rom splines in games~\cite{Dunlop}. The impressive power of splines is demonstrated when procedural landscapes and roads are combined to create entire virtual city models randomly, as can be seen, for instance in the \textit{SceneCity} Blender addon\footnote{SceneCity is a procedural city generation addon for Blender available at: \url{https://www.cgchan.com/}}.
As seen with the CG examples, it is possible to generate paths and surfaces using splines. In addition, Splines can be used to animate objects or virtual cameras along predefined paths \cite{SplineAnimation}. Recently, the research community involved in solving the inverse problem to CG, namely, Computer Vision (CV), increasingly discovers splines for recovering the real camera movement from image sequences. In robotics, related research is often concerned with solving the visual Simultaneous Localization And Mapping (SLAM) problem. However, since cameras usually take pictures at specific points in time, the resulting estimated trajectory is discrete. Imposing constraints on the camera path using splines has shown advantages in handling a common image sensor deficiency called \textit{rolling shutter}, where the image exhibits a pixel warping deformation (informally often called a \textit{Jello Effect}) due to a rowwise exposure~\cite{JelloEffect}. Describing the camera trajectory using continuous B-splines offers the advantage that any exposure time along the path can be interpolated and thus allows for correcting the rolling shutter for both monocular~\cite{SplineBasedFusionMonocular} and stereo (RGB-D)~\cite{SplineBasedFusionStereo} cameras. To predict and plan the future paths for Automated Driving Systems (ADS), while taking into account obstacles and driving comfort, it was shown that Bézier curves are a reliable solution for an optimal trajectory generation~\cite{PathPlanningBezier1, PathPlanningBezier2}.


This short review of interdisciplinary research contains many different types of spline representation, such as Bézier~\cite{DOKKEN199033, PathPlanningBezier1, PathPlanningBezier2}, Catmull–Rom~\cite{Dunlop}, B-Splines~\cite{SplineBasedFusionMonocular,SplineBasedFusionStereo} and Non-uniform Rational B-Spline (NURBS)~\cite{ TesselationSplineSurfaces} surfaces. This shows the relevance of providing conversion methods between the various representations of which one is discussed in this paper.

\subsection{Paper Structure and Contributions}
\label{sec:Contributions}
In this paper, we do not go through the properties and details of Bézier and Catmull–Rom Splines, but rather the aim is to focus on the conversion equations. First, we give a brief explanation of Bézier curves in Sec.~\ref{sec:bezier} as well as a brief explanation of Catmull–Rom Splines in Sec.~\ref{sec:catmull}. 
Then, we go through the conversion equations in Sec.~\ref{sec:conversion}. 
Note that by conversion, we mean conversion of control points of a curve to the control points of another curve, which results in approximately the same curve as the curve before conversion represents\footnote{To encourage reproducibility, the source code of the project is publicly accessible here:
\url{https://github.com/starasteh/bezier_catmullrom/}}. In Sec.~\ref{sec:experimentss}, we validate the equations with some numerical and graphical examples.
We use \textbf{bold-face} notation for vector representation here.
Sec.~\ref{sec:discussion} discusses the conversion equations with regard to the real-world applications and the limitations and proposes some potential future work. Finally, the conclusions are stated in Sec.~\ref{sec:conclusion}.


\section{Bézier Curves}
\label{sec:bezier}

There are two basic ways of defining a curve—in terms of the polygon vertices, and in terms of the polygon sides \cite{101093comjnl1571}. Even though, the latter is the form, which was originally used by Pierre Bézier \cite{bezier1970emploi, bezier1968procede, bezier1968emplo7i}, we consider the former throughout our paper.
There are two reasons for developing a formulation of the
Bézier curve in terms of polygon vertices rather than polygon sides. To begin with, the formulation becomes more elegant. Furthermore, as a general principle, it is better to program in terms of absolute vectors rather than a chain of relative vectors, irrespective of the particular user interface, when transformations such as rotation are to be applied to the vectors, because rounding errors do not have the cumulative effect which sometimes give rise to poor drawings. This can be particularly noticeable when transformations are performed, for reasons of speed, in a small satellite graphics computer \cite{101093comjnl1571}.

\begin{figure*}
  \subfloat[]{
	\begin{minipage}[c][.6\width]{
	   0.3\textwidth}
	   \centering
	   \includegraphics[scale=.34]{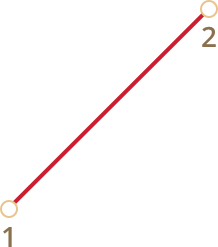}
	\end{minipage}}
 \hfill 	
  \subfloat[]{
	\begin{minipage}[c][.6\width]{
	   0.3\textwidth}
	   \centering
	   \includegraphics[scale=.34]{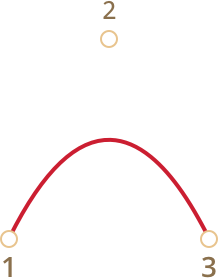}
	\end{minipage}}
 \hfill	
  \subfloat[]{
	\begin{minipage}[c][.6\width]{
	   0.3\textwidth}
	   \centering
	   \includegraphics[scale=.34]{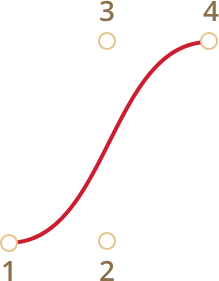}
	\end{minipage}}
\caption{(a) A two-point Bézier curve. (b) A three-point (quadratic) Bézier curve. (c) A four-point (cubic) Bézier curve. Images taken from \cite{name1}.}
\label{fig:multipoints}
\end{figure*}

A parametric Bézier curve is defined by its control points. The curve does not necessarily go through the control points. There may be 2, 3, 4 or more. For instance, Fig.~\ref{fig:multipoints} shows a linear, a quadratic, and cubic Bézier curve, respectively. The figures are drawn using a geometric approach called "de Casteljau's algorithm". It is the most common approach for drawing Bézier curves in CAGD. For more details on the algorithm, see \cite{farinmorgan}.

A cubic Bézier curve can be written in a matrix form by expanding the analytic definition of the curve into its Bernstein polynomial coefficients, and then writing these coefficients in a matrix form using the polynomial power basis \cite{bbezier}. The Bernstein polynomials\footnote{For more information on the Bernstein polynomials, see \cite{bernstein, farinmorgan, bernsteingerman}.} of degree $n$ are defined by Eq.~\ref{eq:bernstein1}. For more details on Bézier curves, see~\cite{name3}.

\begin{dmath}
B_{i,n}(t)={n \choose i}t^i(1-t)^{n-i}
\label{eq:bernstein1}
\end{dmath}
A Bézier curve uses Bernstein polynomials as basis. A cubic (degree 3 or order 4) is represented by the following,
\begin{dmath}
\label{eq:bezier23}
\text{Bezier}(t) =
\sum_{i=0}^{3}{\mathbf{P_iB_i(t)}}=(1-t)^3\mathbf{P_0}+3t(1-t)^2\mathbf{P_1}+3t^2(1-t)\mathbf{P_2}+t^3\mathbf{P_3}
\end{dmath}
\begin{dmath}
=\begin{bmatrix}
1&t&t^2&t^3
\end{bmatrix}\cdot
\begin{bmatrix}
1&0&0&0 \\
-3&3&0&0 \\
3&-6&3&0 \\
-1&3&-3&1
\end{bmatrix}\cdot
\begin{bmatrix}
\mathbf{P_0}\\
\mathbf{P_1}\\
\mathbf{P_2}\\
\mathbf{P_3 }\label{eq:bezier}
\end{bmatrix}.
\end{dmath}


\section{Catmull–Rom Splines}
\label{sec:catmull}

\begin{figure}
\centering
\includegraphics[scale=.3]{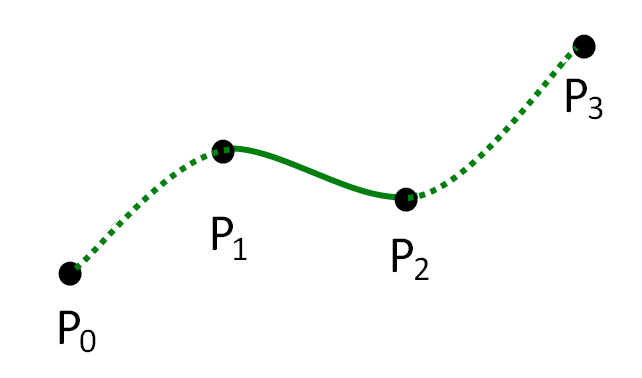}
\caption{Catmull–Rom spline interpolation with four points. Image taken from \cite{catmullnew}.}
\label{fig:catmull}
\end{figure}

Catmull–Rom splines are a family of cubic interpolating splines formulated such that the tangent at each point $\mathbf {P} _{i}$ is calculated using the previous and next point on the spline \cite{twigg2003catmull}.
Unlike a Bézier curve, a (Centripetal) Catmull–Rom spline is defined for only 4 control points (see Fig.~\ref{fig:catmull}), i.e., a single Catmull–Rom segment is cubic. 
As it is an interpolating spline, the curve goes, through its control points, $ \mathbf {P} _{0},\mathbf {P} _{1},\mathbf {P} _{2},\mathbf {P} _{3}$, and it is only drawn from $\mathbf {P} _{1}$ to $ \mathbf {P} _{2}$  (Fig.~\ref{fig:catmull}).
The curve is named after Edwin Catmull and Raphael Rom. The principal advantage of this technique is that the points along the original set of points also make up the control points for the spline curve\cite{edwincatmull}.

Catmull–Rom splines are based on the concept of "tension": the higher the tensions, the shorter the tangents at the departure and arrival points. 
It affects how sharply the curve bends at the (interpolated)
control points \cite{twigg2003catmull}.
When tension is set to 1 (as it often is), the resulting segments between control points will be straight lines. 
The basic Catmull–Rom curve arrives and departs with tangents equal to half the distance between the two adjacent points. 

The matrix form of a Catmull–Rom\footnote{$\mathbf{P_i'}, i\in \lbrace0,1,2,3\rbrace$ show the Catmull–Rom spline control points.} spline is shown in Eq.~\ref{eq:catmull},

\begin{dmath}
\text{CatmullRom}(t) =\frac{1}{2}\cdot \begin{bmatrix}
1&t&t^2&t^3
\end{bmatrix}\cdot
\begin{bmatrix}
0&2&0&0 \\
-\tau&0&\tau&0 \\
2\tau&\tau-6&-2(\tau-3)&-\tau \\
-\tau&4-\tau&\tau-4&\tau
\end{bmatrix}.
\begin{bmatrix}
\mathbf{P'_0}\\
\mathbf{P'_1}\\
\mathbf{P'_2}\\
\mathbf{P'_3} \label{eq:catmull}
\end{bmatrix}
\end{dmath}
where $\tau$ is the tension factor\footnote{In many references, you may find the matrix form of the Catmull–Rom splines as, \begin{dmath}
\text{CatmullRom}(t) =\frac{1}{2}\cdot \begin{bmatrix}
1&t&t^2&t^3
\end{bmatrix}\cdot
\begin{bmatrix}
0&2&0&0 \\
-1&0&1&0 \\
2&-5&4&-1 \\
-1&3&-3&1
\end{bmatrix}.
\begin{bmatrix}
\mathbf{P'_0}\\
\mathbf{P'_1}\\
\mathbf{P'_2}\\
\mathbf{P'_3}  
\end{bmatrix}
\end{dmath} which is the case when $\tau = 1$.}. See \cite{catmullproof} for the proof of Eq.~\ref{eq:catmull} with $\tau = 1$.


\section{Conversion}
\label{sec:conversion}

We will use only cubic splines for both Catmull–Rom and Bézier throughout this paper, as the Catmull–Rom splines are only defined with four control points. It is always possible to concatenate multiple splines to create the desired curve. A series of unique cubic polynomials are fitted between each of the data points, with the stipulation that the curve obtained be continuous and appear smooth \cite{mckinley1998cubic}.
As a matter of fact, in industrial applications, it is often found that a particular curve segment is not sufficiently powerful or flexible (i.e., it does not have sufficient degrees of freedom) to adopt a desired shape \cite{101093comjnl1571}.

Comparing Eq.~\ref{eq:bezier} and Eq.~\ref{eq:catmull}, we wish to find the transformation matrix $A$ in order to convert from Catmull–Rom to Bézier,

\begin{dmath}
\begin{split}
{}
\begin{bmatrix}
1&t&t^2&t^3
\end{bmatrix}\cdot\frac{1}{2}\cdot
\begin{bmatrix}
0&2&0&0 \\
-\tau&0&\tau&0 \\
2\tau&\tau-6&-2(\tau-3)&-\tau \\
-\tau&4-\tau&\tau-4&\tau
\end{bmatrix}.
\begin{bmatrix}
\mathbf{P'_0}\\
\mathbf{P'_1}\\
\mathbf{P'_2}\\
\mathbf{P'_3} 
\end{bmatrix}\\
=
\begin{bmatrix}
1&t&t^2&t^3
\end{bmatrix}\cdot
\begin{bmatrix}
1&0&0&0 \\
-3&3&0&0 \\
3&-6&3&0 \\
-1&3&-3&1
\end{bmatrix}\cdot A\cdot
\begin{bmatrix}
\mathbf{P_0}\\
\mathbf{P_1}\\
\mathbf{P_2}\\
\mathbf{P_3} \label{eq:5}
\end{bmatrix}.
\end{split}
\end{dmath}
Consequently, we have,
\begin{dmath}
\begin{bmatrix}
1&t&t^2&t^3
\end{bmatrix}\cdot
M' \cdot
\begin{bmatrix}
\mathbf{P'_0}\\
\mathbf{P'_1}\\
\mathbf{P'_2}\\
\mathbf{P'_3} 
\end{bmatrix} =
\begin{bmatrix}
1&t&t^2&t^3
\end{bmatrix}\cdot
M \cdot
\begin{bmatrix}
\mathbf{P_0}\\
\mathbf{P_1}\\
\mathbf{P_2}\\
\mathbf{P_3} 
\end{bmatrix}
\end{dmath}
with $M$ and $M'$ being the following matrices,

\begin{dmath}
M'=\frac{1}{2}\cdot
\begin{bmatrix}
0&2&0&0 \\
-\tau&0&\tau&0 \\
2\tau&\tau-6&-2(\tau-3)&-\tau \\
-\tau&4-\tau&\tau-4&\tau
\end{bmatrix}\\
\end{dmath}

\begin{dmath}
M= \begin{bmatrix}
1&0&0&0 \\
-3&3&0&0 \\
3&-6&3&0 \\
-1&3&-3&1
\end{bmatrix}\cdot A
\end{dmath}

The difference is somewhere in the $M$ and $M'$ matrices, since $t$ and the coordinate values are identical. Therefore, we should solve the matrix equation in Eq.~\ref{eq:6}.

\begin{dmath}
\frac{1}{2}\cdot
\begin{bmatrix}
0&2&0&0 \\
-\tau&0&\tau&0 \\
2\tau&\tau-6&-2(\tau-3)&-\tau \\
-\tau&4-\tau&\tau-4&\tau
\end{bmatrix}=
\begin{bmatrix}
1&0&0&0 \\
-3&3&0&0 \\
3&-6&3&0 \\
-1&3&-3&1
\end{bmatrix}\cdot A \label{eq:6}
\end{dmath}
We left-multiply both sides by the inverse of the Bézier matrix, to get rid of the Bézier matrix on the right side of the equals sign,
\begin{dmath}
\begin{split}
{}
\begin{bmatrix}
1&0&0&0 \\
-3&3&0&0 \\
3&-6&3&0 \\
-1&3&-3&1
\end{bmatrix}^{-1}
\cdot\frac{1}{2}\cdot
\begin{bmatrix}
0&2&0&0 \\
-\tau&0&\tau&0 \\
2\tau&\tau-6&-2(\tau-3)&-\tau \\
-\tau&4-\tau&\tau-4&\tau 
\end{bmatrix}\\ 
=
\begin{bmatrix}
1&0&0&0 \\
-3&3&0&0 \\
3&-6&3&0 \\
-1&3&-3&1
\end{bmatrix}^{-1}\cdot
\begin{bmatrix}
1&0&0&0 \\
-3&3&0&0 \\
3&-6&3&0 \\
-1&3&-3&1
\end{bmatrix}\cdot A=I\cdot A=A, \label{eq:7}
\end{split}
\end{dmath}
which brings us to Eq.~\ref{eq:8}.

\begin{dmath}
\frac{1}{6}\cdot\begin{bmatrix}
0&6&0&0\\
-\tau&6&\tau&0\\
0&\tau&0&-\tau\\
0&0&6&0
\end{bmatrix}=A. \label{eq:8}
\end{dmath}
Multiplying this \textit{A} with our coordinates will give us a proper Bézier matrix expression again, as in Eq.~\ref{eq:9} and Eq.~\ref{eq:10}.

\begin{dmath}
\begin{bmatrix}\label{eq:9}
1&t&t^2&t^3
\end{bmatrix}
\begin{bmatrix}
1&0&0&0 \\
-3&3&0&0 \\
3&-6&3&0 \\
-1&3&-3&1
\end{bmatrix}\cdot
\frac{1}{6}\cdot\begin{bmatrix}
0&6&0&0\\
-\tau&6&\tau&0\\
0&\tau&0&-\tau\\
0&0&6&0
\end{bmatrix}\cdot
\begin{bmatrix}
\mathbf{P_0}\\
\mathbf{P_1}\\
\mathbf{P_2}\\
\mathbf{P_3} 
\end{bmatrix}
\end{dmath}

\begin{dmath}
=\begin{bmatrix}
1&t&t^2&t^3
\end{bmatrix}
\begin{bmatrix}
1&0&0&0 \\
-3&3&0&0 \\
3&-6&3&0 \\
-1&3&-3&1
\end{bmatrix}\cdot
\begin{bmatrix}
\mathbf{P_1}\\
\mathbf{P_1}+\displaystyle{\frac{\mathbf{P_2}-\mathbf{P_0}}{6\cdot\tau}}\\
\mathbf{P_2}-\displaystyle{\frac{\mathbf{P_3}-\mathbf{P_1}}{6\cdot\tau}}\\
\mathbf{P_2} \label{eq:10}
\end{bmatrix}.
\end{dmath}

Thus, a Catmull–Rom to Bézier conversion, based on coordinates, requires turning the Catmull–Rom coordinates on the left into the Bézier coordinates on the right (with $\tau$ being our tension factor), according to Eq.~\ref{eq:11}.

\begin{align}
\begin{bmatrix}
\mathbf{P'_0}\\
\mathbf{P'_1}\\
\mathbf{P'_2}\\
\mathbf{P'_3} 
\end{bmatrix}_{CatmullRom}\Rightarrow
\begin{bmatrix}
\mathbf{P_1}\\
\mathbf{P_1}+\displaystyle{\frac{\mathbf{P_2}-\mathbf{P_0}}{6\cdot\tau}}\\
\mathbf{P_2}-\displaystyle{\frac{\mathbf{P_3}-\mathbf{P_1}}{6\cdot\tau}}\\
\mathbf{P_2} \label{eq:11}
\end{bmatrix}_{Bezier}
\end{align}
In addition, in the same way, a Bézier to Catmull–Rom conversion instead requires turning the Bézier coordinates on the left into the Catmull–Rom coordinates on the right. Note that, there is no tension factor this time, because Bézier curves do not have any. Converting from Bézier to Catmull–Rom is simply a default-tension Catmull–Rom curve, based on Eq.~\ref{eq:12}.
\begin{align}
\begin{bmatrix}
\mathbf{P_0}\\
\mathbf{P_1}\\
\mathbf{P_2}\\
\mathbf{P_3 }
\end{bmatrix}_{Bezier}\Rightarrow
\begin{bmatrix}
\mathbf{P'_3}+6\cdot(\mathbf{P'_0}-\mathbf{P'_1})\\
\mathbf{P'_0}\\
\mathbf{P'_3}\\
\mathbf{P'_0}+6\cdot(\mathbf{P'_3}-\mathbf{P'_2}) \label{eq:12}
\end{bmatrix}_{CatmullRom}
\end{align}

\section{Experiments and Results}
\label{sec:experimentss}

In this section, we use sets of exemplary control points in order to validate the transformation equations.

\subsection{Numerical Analysis}
Having the fact that we only consider some random values here, this part cannot be considered as a mathematical proof. However, it gives a sound overview of how the conversion equations (Eq.~\ref{eq:11} and Eq.~\ref{eq:12}) work.

Looking at the equations, it is trivial that the conversion equations are valid for any number of dimensions. In this part, we explain our examples with 3-dimensional (3D) control points.

In the following, first, we validate the conversion from cubic Bézier curves to cubic Catmull–Rom splines by experimenting two random timepoints. And in the second part, we do the reverse and validate the conversion from cubic Catmull–Rom splines to cubic Bézier curves by experimenting two other random timepoints.


\subsubsection{Conversion from Bézier to Catmull–Rom}
Let us have a cubice Bézier curve with the following 3D control points,
\begin{eqnarray*}
\mathbf{P_0} &=& (1.0, 1.0, 1.0),\\
\mathbf{P_1} &=& (2.0, 2.0, 2.0),\\
\mathbf{P_2} &=& (3.0, 3.0, 3.0),\\
\mathbf{P_3} &=&  (4.0, 4.0, 4.0).
\end{eqnarray*}
Using Eq.~\ref{eq:12}, we get the following control points for the Catmull–Rom spline with the tension factor $\tau =1$,
\begin{eqnarray*}
\mathbf{P'_0} &=& (-2.0, -2.0, -2.0),\\
\mathbf{P'_1} &=& (1.0, 1.0, 1.0),\\
\mathbf{P'_2} &=& (4.0, 4.0, 4.0),\\
\mathbf{P'_3} &=&  (7.0, 7.0, 7.0).
\end{eqnarray*}
Now, we replace the set of control data points in Eq.~\ref{eq:bezier} and with a random choice of $t= 0.2$, we get the following,

\begin{dmath}
\text{Bezier}(0.2)=
\begin{bmatrix}
1&0.2&0.04&0.008
\end{bmatrix}\cdot
\begin{bmatrix}
1&0&0&0 \\
-3&3&0&0 \\
3&-6&3&0 \\
-1&3&-3&1
\end{bmatrix}\cdot
\begin{bmatrix}
1.0\\
2.0\\
3.0\\
4.0\end{bmatrix} = 1.600.
\end{dmath}
The obtained value should be the same, when using Eq.~\ref{eq:catmull} with $t= 0.2$,
\begin{dmath}
{1.600=\text{CatmullRom}(0.2)}= 
\frac{1}{2} \cdot \begin{bmatrix}
1&0.2&0.04&0.008
\end{bmatrix}\cdot
\begin{bmatrix}
0&2&0&0 \\
-1&0&1&0 \\
2&-5&4&-1 \\
-1&3&-3&1
\end{bmatrix}.
\begin{bmatrix}
\mathbf{P'_0}\\
\mathbf{P'_1}\\
\mathbf{P'_2}\\
\mathbf{P'_3}  
\end{bmatrix}.
\end{dmath}
Solving the matrix equation, we get the following control points,
\begin{eqnarray*}
\mathbf{P'_0} &=& (-2.0, -2.0, -2.0),\\
\mathbf{P'_1} &=& (1.0, 1.0, 1.0),\\
\mathbf{P'_2} &=& (4.0, 4.0, 4.0),\\
\mathbf{P'_3} &=&  (7.0, 7.0, 7.0).
\end{eqnarray*}
which are equal to the results of using  Eq.~\ref{eq:12}. \qed
 
 \paragraph{Another timepoint}
Now, to show the equation is valid for any timepoint, we choose another random time $t= 0.64$ and follow the same procedure as above for the same control points,
\begin{dmath}
\text{Bezier}(0.64)=
\begin{bmatrix}
1&0.64&0.4096&0.262144
\end{bmatrix}\cdot
\begin{bmatrix}
1&0&0&0 \\
-3&3&0&0 \\
3&-6&3&0 \\
-1&3&-3&1
\end{bmatrix}\cdot
\begin{bmatrix}
1.0\\
2.0\\
3.0\\
4.0\end{bmatrix} = 2.920
\end{dmath}

\begin{dmath}
2.920=\text{CatmullRom}(0.64) =\frac{1}{2}\cdot \begin{bmatrix}
1&0.64&0.4096&0.262144
\end{bmatrix}\cdot
\begin{bmatrix}
0&2&0&0 \\
-1&0&1&0 \\
2&-5&4&-1 \\
-1&3&-3&1
\end{bmatrix}.
\begin{bmatrix}
\mathbf{P'_0}\\
\mathbf{P'_1}\\
\mathbf{P'_2}\\
\mathbf{P'_3} 
\end{bmatrix}.
\end{dmath}
Solving the matrix equation, we again get the following values,
\begin{eqnarray*}
\mathbf{P'_0} &=& (-2.0, -2.0, -2.0)\\
\mathbf{P'_1} &=& (1.0, 1.0, 1.0)\\
\mathbf{P'_2} &=& (4.0, 4.0, 4.0)\\
\mathbf{P'_3} &=&  (7.0, 7.0, 7.0).
\end{eqnarray*} 
which are again equal to the results of using  Eq.~\ref{eq:12}. \qed

\subsubsection{Conversion from Catmull–Rom to Bézier}

In this part, we do the reverse and start with a cubic Catmull–Rom spline with the following 3D control points,
\begin{eqnarray*}
\mathbf{P'_0} &=& (1.0, 1.0, 1.0),\\
\mathbf{P'_1} &=& (2.0, 2.0, 2.0),\\
\mathbf{P'_2} &=& (3.0, 3.0, 3.0),\\
\mathbf{P'_3} &=&  (4.0, 4.0, 4.0).
\end{eqnarray*}
Using Eq.~\ref{eq:11}, we get the following control points for the Bézier curve with the tension factor $\tau =1$,
\begin{eqnarray*}
\mathbf{P_0} &=& (2.0, 2.0, 2.0),\\
\mathbf{P_1} &=& (1.0, 1.0, 1.0),\\
\mathbf{P_2} &=& (2.0, 2.0, 2.0),\\
\mathbf{P_3} &=&  (3.0, 3.0, 3.0).
\end{eqnarray*}
Replacing the set of control data points in Eq.~\ref{eq:catmull} and with a random choice of $t= 0.45$, we get the following,

\begin{dmath}
{\text{CatmullRom}(0.45)}= 
\frac{1}{2} \cdot \begin{bmatrix}
1&0.45&0.2025&0.0911
\end{bmatrix}\cdot
\begin{bmatrix}
0&2&0&0 \\
-1&0&1&0 \\
2&-5&4&-1 \\
-1&3&-3&1
\end{bmatrix}.
\begin{bmatrix}
1.0\\
2.0\\
3.0\\
4.0 
\end{bmatrix} = 2.5411.
\end{dmath}
The obtained value should be the same, when using Eq.~\ref{eq:bezier} with $t= 0.45$,

\begin{dmath}
2.5411=\text{Bezier}(0.45)=
\begin{bmatrix}
1&0.45&0.2025&0.0911
\end{bmatrix}\cdot
\begin{bmatrix}
1&0&0&0 \\
-3&3&0&0 \\
3&-6&3&0 \\
-1&3&-3&1
\end{bmatrix}\cdot
\begin{bmatrix}
\mathbf{P_0}\\
\mathbf{P_1}\\
\mathbf{P_2}\\
\mathbf{P_3}
\end{bmatrix}.
\end{dmath}
Solving the matrix equation, we again get the following values,
\begin{eqnarray*}
\mathbf{P_0} &=& (2.0, 2.0, 2.0),\\
\mathbf{P_1} &=& (1.0, 1.0, 1.0),\\
\mathbf{P_2} &=& (2.0, 2.0, 2.0),\\
\mathbf{P_3} &=&  (3.0, 3.0, 3.0),
\end{eqnarray*} 
which are equal to the results of using  Eq.~\ref{eq:11}. \qed

\paragraph{Another timepoint}
Now, to show the equation is valid for any timepoint, we choose another random time $t= 0.32$ and follow the same procedure as above for the same control points,
\begin{dmath}
{\text{CatmullRom}(0.32)}= 
\frac{1}{2} \cdot \begin{bmatrix}
1&0.32&0.1024&0.0327
\end{bmatrix}\cdot
\begin{bmatrix}
0&2&0&0 \\
-1&0&1&0 \\
2&-5&4&-1 \\
-1&3&-3&1
\end{bmatrix}.
\begin{bmatrix}
1.0\\
2.0\\
3.0\\
4.0 
\end{bmatrix} = 2.3527.
\end{dmath}

\begin{dmath}
2.3527=\text{Bezier}(32)=
\begin{bmatrix}
1&0.32&0.1024&0.0327
\end{bmatrix}\cdot
\begin{bmatrix}
1&0&0&0 \\
-3&3&0&0 \\
3&-6&3&0 \\
-1&3&-3&1
\end{bmatrix}\cdot
\begin{bmatrix}
\mathbf{P_0}\\
\mathbf{P_1}\\
\mathbf{P_2}\\
\mathbf{P_3}
\end{bmatrix}.
\end{dmath}
Solving the matrix equation, we again get the following values,
\begin{eqnarray*}
\mathbf{P_0} &=& (2.0, 2.0, 2.0),\\
\mathbf{P_1} &=& (1.0, 1.0, 1.0),\\
\mathbf{P_2} &=& (2.0, 2.0, 2.0),\\
\mathbf{P_3} &=&  (3.0, 3.0, 3.0),
\end{eqnarray*} 
which are again equal to the results of using  Eq.~\ref{eq:11}. \qed


\subsection{Graphical Fitting}

\begin{figure*}[ht]
\begin{subfigure}{.49\linewidth}
\caption{}
\label{fig:beziertocatmulla}
\begin{tikzpicture}
\begin{axis}[
                 grid=both,
                 grid style={solid,gray!30!white},
                 axis lines=middle,,]
\addplot[blue,thick] table [x=Step, y=Value, col sep=comma] {b2ca.csv};
\end{axis}
\end{tikzpicture}
\end{subfigure}%
\begin{subfigure}{.49\linewidth}
\caption{}
\label{fig:beziertocatmullb}
\begin{tikzpicture}
\begin{axis}[
                 grid=both,
                 grid style={solid,gray!30!white},
                 axis lines=middle,,]
\addplot[red,thick] table [x=Step, y=Value, col sep=comma] {b2cb.csv};
\end{axis}
\end{tikzpicture}
\end{subfigure}
\caption{2D curves converted from Bézier to Catmull–Rom, drawn for 1000 time steps. (a) The original Bézier curve, which is drawn using the following randomly-generated 2D control points: $\mathbf{P_0} = (70.650, 40.045),\: \mathbf{P_1} = (65.354, 15.054),\: \mathbf{P_2} = (60.253, 100.754),\: \mathbf{P_3} = (100.234, 95.287)$. (b) The calculated Catmull–Rom curve with $\tau=1$, which is drawn using the following control points, which are the converted control point of the original Bézier curve using Eq.~\ref{eq:12}: $\mathbf{P'_0} = (130.010, 245.233),\: \mathbf{P'_1} = (70.650, 40.045),\: \mathbf{P'_2} = (100.234, 95.287),\: \mathbf{P'_3} = (310.536, 7.243)$.}
\label{fig:beziertocatmull}
\end{figure*}
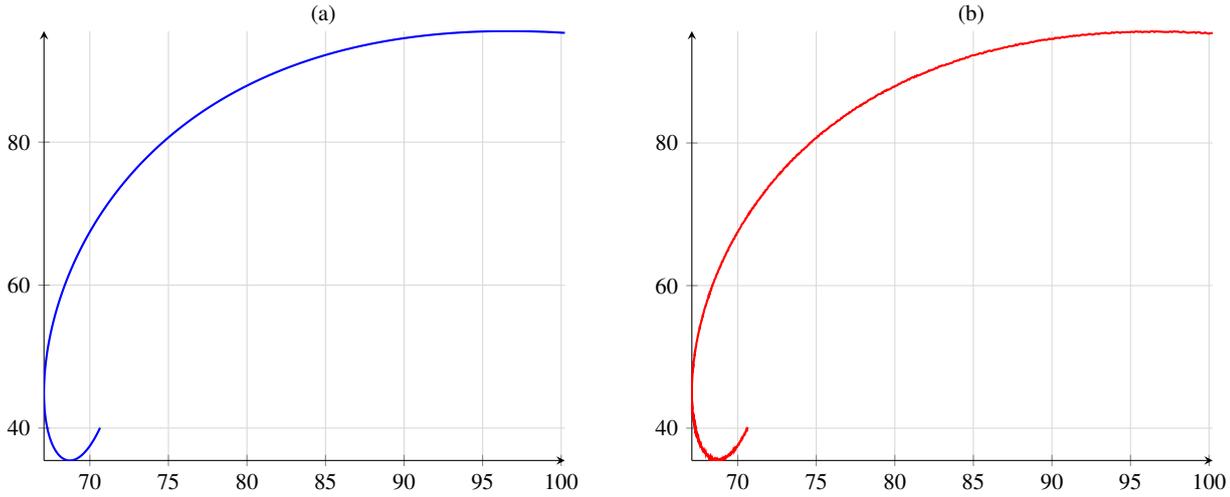

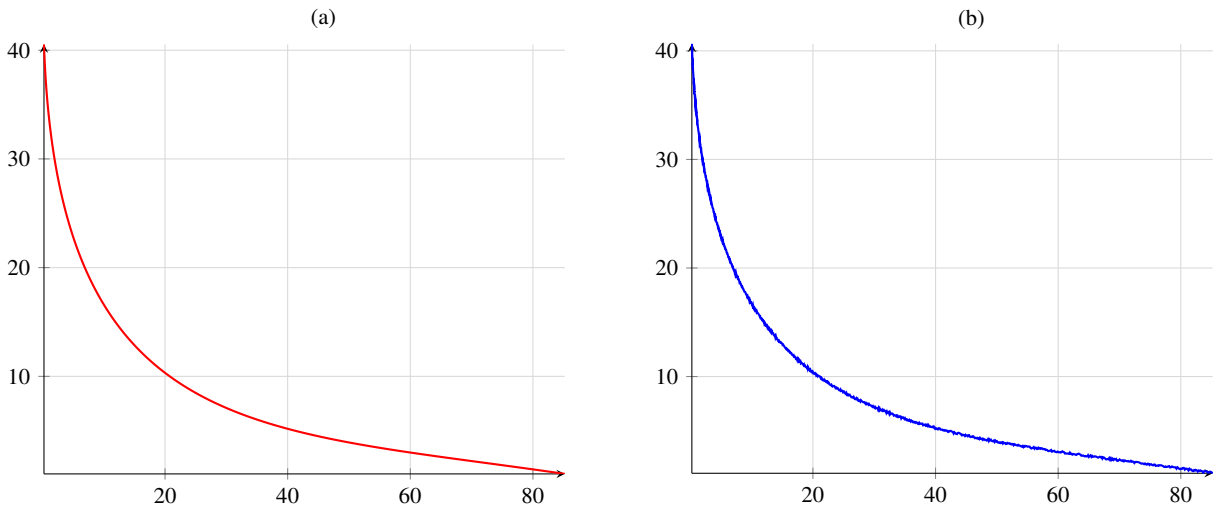
\begin{figure*}[ht]
\begin{subfigure}{.49\linewidth}
\caption{}
\label{fig:catmulltobeziera}
\begin{tikzpicture}
\begin{axis}[
                 grid=both,
                 grid style={solid,gray!30!white},
                 axis lines=middle,,]
\addplot[red,thick] table [x=Step, y=Value, col sep=comma] {c2ba.csv};
\end{axis}
\end{tikzpicture}
\end{subfigure}%
\begin{subfigure}{.49\linewidth}
\caption{}
\label{fig:catmulltobezierb}
\begin{tikzpicture}
\begin{axis}[
                 grid=both,
                 grid style={solid,gray!30!white},
                 axis lines=middle,,]
\addplot[blue,thick] table [x=Step, y=Value, col sep=comma] {c2bb.csv};
\end{axis}
\end{tikzpicture}
\end{subfigure}
\caption{2D curves converted from Catmull–Rom to Bézier, drawn for 1000 time steps. (a) The original Catmull–Rom curve with $\tau=1$, which is drawn using the following randomly-generated 2D control points: $\mathbf{P'_0} =  (72.022, 219.863),\: \mathbf{P'_1} = (0.257, 40.527),\: \mathbf{P'_2} =  (85.204, 1.025),\: \mathbf{P'_3} = (307.331, 15.189)$. (b) The calculated Bézier curve, which is drawn using the following control points, which are the converted control point of the original Catmull–Rom curve using Eq.~\ref{eq:11}: $\mathbf{P_0} = (0.257, 40.527),\: \mathbf{P_1} = (2.454, 4.054),\: \mathbf{P_2} = (34.025, 5.248),\: \mathbf{P_3} = (85.204, 1.025)$.}
\label{fig:catmulltobezier}
\end{figure*}

In this part, we aim at showing the validity of the conversion equations (Eq.~\ref{eq:11} and Eq.~\ref{eq:12}) by illustrating graphical examples.
For the ease of illustration, we assume 2-dimensional (2D) curves in this part.

Figure~\ref{fig:beziertocatmulla} shows a cubic Bézier curve which is drawn for 1000 timepoints ($t \in [0,1]$ with a step size of 0.001). Furthermore, the aforementioned figure is drawn using the following randomly-generated 2D control points, 
\begin{eqnarray*}
\mathbf{P_0} &=&  (70.650, 40.045),\\
\mathbf{P_1} &=&  (65.354, 15.054),\\
\mathbf{P_2} &=&  (60.253, 100.754),\\
\mathbf{P_3} &=&  (100.234, 95.287).
\end{eqnarray*} 
Using Eq.~\ref{eq:12}, we convert the above Bézier control points to the respective control points of a Catmull–Rom cubic spline with the tension factor $\tau =1$,
\begin{eqnarray*}
\mathbf{P'_0} &=&  (130.010, 245.233),\\
\mathbf{P'_1} &=&  (70.650, 40.045),\\
\mathbf{P'_2} &=&  (100.234, 95.287),\\
\mathbf{P'_3} &=&  (310.536, 7.243),
\end{eqnarray*}
and sketch the corresponding cubic Catmull–Rom spline of these new control points for the same 1000 timepoints ($t \in [0,1]$ with a step size of 0.001).
Fig.~\ref{fig:beziertocatmullb} illustrates the resulting curve, which is almost the same as Fig.~\ref{fig:beziertocatmulla}.

In a similar way, Fig.~\ref{fig:catmulltobezier} shows the reverse conversion from Catmull–Rom to Bézier using Eq.~\ref{eq:11} and the following randomly-generated 2D control points and the resulting converted control points, 
\begin{eqnarray*}
\mathbf{P'_0} &=&  (72.022, 219.863),\\
\mathbf{P'_1} &=&  (0.257, 40.527),\\
\mathbf{P'_2} &=&  (85.204, 1.025),\\
\mathbf{P'_3} &=&  (307.331, 15.189),
\end{eqnarray*} 

\begin{eqnarray*}
\mathbf{P_0} &=&  (0.257, 40.527),\\
\mathbf{P_1} &=&  (2.454, 4.054),\\
\mathbf{P_2} &=&  (34.025, 5.248),\\
\mathbf{P_3} &=&  (85.204, 1.025).
\end{eqnarray*}


\section{Discussion}
\label{sec:discussion}

The method proposed in this work allows for converting control points between Bézier and Catmull–Rom definitions. This conversion can be applied in situations, where one of these mathematical representations is not supported in specific software implementations and thus allows to use data, such as 2D drawings, 3D camera trajectories or surfaces across individual software pipelines. For instance, the open source 3D creation suite Blender does not support Catmull–Rom curve primitives\footnote{See the supported curve types in Blender here: \url{https://docs.blender.org/manual/en/latest/modeling/curves/primitives.html}} but it does support Bézier curves. Therefore, one of the many applications of the proposed method in this paper would be to convert the control points of a Catmull–Rom curve to the Bézier representation in order to be able to draw the curve in Blender.

This work has also some limitations. The proposed method is solely valid for cubic curve segments, i.e., curves which are defined by four control points. However, Bézier curves may be defined with an arbitrary number of control points and consequently, contain a higher-degree polynomial. The advantage of lower-degree polynomials is less computations during both the evaluation of the curve and the calculation of derivatives (comparing the definitions in Eq.~\ref{eq:bernstein1} and Eq.~\ref{eq:bezier23}, the number of summations/loops depends on the degree of the curve). Therefore, the higher the degree, the higher the number of evaluations. In order to use our proposed method with higher-degree curves, additional techniques such as \textit{Degree Reduction}~\cite{BezierDegreeReduction} or \textit{Curve Subdivision} using De Casteljau's algorithm~\cite{DeCasteljau} need to be applied to the curve prior to type conversion. Moreover, B-Splines might be utilized in many applications instead of Bézier splines, due to the fact that they could provide finer local shape control by introducing a \textit{knot vector}. Even though there is a conversion method between B-Splines and Bézier splines~\cite{Prautzsch2002Bezier}, the authors are not aware of a method to directly convert between B-Splines and Catmull–Rom splines. Such generalizations to other spline representations are expected to be formulated in future research.


\section{Conclusion}
\label{sec:conclusion}

In this paper, we focused on conversion of control points of a cubic Bézier curve to those of a Catmull–Rom curve and vice-versa.
It was shown that, to do so, according to Eq.~\ref{eq:11} and Eq.~\ref{eq:12}, we merely need basic linear transformations of the positions of the control points. Moreover, we illustrated that the equations are valid for any timepoint and for control points with arbitrary number of dimensions.

\begin{acknowledgements}

This work was supported by Siemens Mobility GmbH, Erlangen, Germany.

\end{acknowledgements}

\section*{Conflict of interest}

All authors declare that they have no conflict of interest.

\bibliographystyle{spmpsci}      
\bibliography{template}   

\end{document}